# Ionic Charge Imbalance in Perovskite Solar Cells

Dhyana Sivadas, Abhimanyu Singareddy, Chittiboina Ganga Vinod and Pradeep R. Nair
Dept. of Electrical Engineering, Indian Institute of Technology Bombay, Mumbai, India

Ion migration in perovskite solar cells is usually analyzed and understood in terms of charge neutrality condition. However, several recent reports indicate possibility of ionic imbalance in the active layer due to external ion migration and/or chemical reactions. In this context, here we explore the influence of ionic charge neutrality on the performance of perovskite solar cells. Our results indicate that ionic imbalance leads to an asymmetry in the device electrostatics, which have interesting implications on the impact of material/interface degradation, hysteresis, and finally on the long-term stability and influence of optimal device architecture (NIP vs. PIN).

**INTRODUCTION:**

Ion migration is considered as a contributing factor to several phenomena like hysteresis and long-term stability in Perovskite solar cells (PSCs)[1–4]. Typically, the ions under consideration originate from the dissociation of (or defects in) perovskite material itself [5–8]. Accordingly, the active layer contains an equivalent amount of positive and negative ions – rather the net or the integrated ionic charge is zero. Indeed, such an assumption of ionic charge neutrality is integral to almost all modeling efforts attempting to unravel hysteresis and performance degradation[5,9]. However, such an assumption of ionic charge neutrality might not be valid as indicated in several recent experimental observations. For example, perovskite has a low diffusion barrier for elemental migration and the transport layers (TLs) are often permeable to ionic transport[10]. This enables movement of mobile ions from the perovskite to the outer layers and the migration of external ions (metal ions from contact or ions from TL) to the perovskite[11–18]. Stressors like temperature, moisture, and light accelerate the formation of mobile ions and increase the diffusivity of ions across the material layers[11,17,19] – all of these could lead to an imbalance of ionic charge in the perovskite active region.

The imbalance of ions inside the active layer could lead to performance degradation in the device[11,13]. Hence, it is critical to comprehend the implications of ionic imbalance for predicting the long-term stability of PSCs. A detailed understanding about the underlying physics of the same is still lacking in literature. Some unanswered questions are: (i) Are there any notable differences in the device electrostatics due to inter layer ion migration? (ii) If so, what might be the impact under significant interface/bulk recombination? (iii) How does such a device perform under practical field condition, for example, at elevated temperatures?

Seeking to gain more insights, we model and analyze the impact of unbalanced ions on the performance of perovskite solar cells. With the help of numerical simulations, we find that the device with ionic imbalance has distinct electrostatics as compared to devices with ionic charge neutrality. Curiously, ionic imbalance in the absorber manifests as an apparent unintentional doping which leads to an asymmetrical junction formation in the device. This electrostatic asymmetry results in distinct device characteristics depending on the spatial location of material degradation and illumination side (i.e., NIP vs PIN schemes). Besides, we examine the influence of ionic imbalance on hysteresis and temperature coefficient of perovskite solar cells.

**MODEL SYSTEM:**

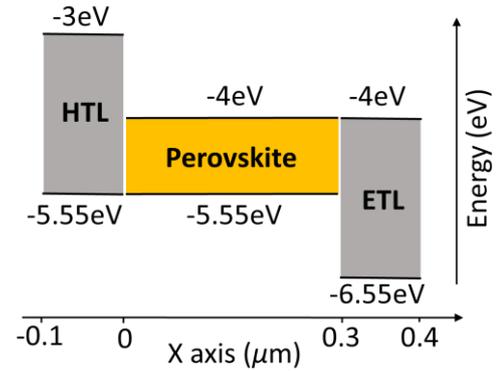

**FIG. 1.** Schematic diagram of Perovskite solar cell with energy alignment of materials.

The schematic of the device and energy levels of the materials are shown in Fig. 1. The device consists of a PIN structure with a p doped-hole transport layer (HTL), an intrinsic-perovskite layer, and an n doped-electron transport layer (ETL). Here, the conduction band of the perovskite is aligned with the conduction band of the ETL and the valance band of the perovskite is aligned with the valance band of the HTL (effects of band offsets are considered later). Irrespective of the specifics of parameters related to band alignment, this system can serve as a model to study the influence of ionic charge imbalance as described below.

We use self-consistent drift diffusion simulations[20] to understand the behavior of ions inside the system. The model consists of the Poisson's equation (eq. 1), continuity equation for electrons (eq. 2) and continuity equation for holes (eq. 3).

$$\frac{d^2V}{dx^2} = -\frac{q}{\varepsilon}\left(p - n + N_{dop} + N_{I,p} - N_{I,n}\right) \quad (1)$$

$$\frac{1}{q}\frac{\partial J_n(x)}{\partial x} + G - R = 0 \quad (2)$$

$$-\frac{1}{q}\frac{\partial J_p(x)}{\partial x} + G - R = 0 \quad (3)$$

Here $V$ is the potential, $x$ is the spatial coordinate, $q$ is the elementary charge and $\varepsilon$ is the dielectric permittivity. Electron and hole carrier densities are denoted by $n$ and $p$ respectively. The parameter $N_{dop}$ represents the doping concentration of materials (see section I, Suppl. Mat. for material parameters). $N_{I,p}$, $N_{I,n}$ are the positive and the negative ion concentrations in the perovskite layer. Further, $J_n$ and $J_p$ indicate the electron current and the hole current. Also, we consider a photo-generation rate, $G$ inside the perovskite layer. The recombination rate ($R$) is modelled by SRH, Auger and radiative[21,22] (see section II, Suppl. Mat.). Moreover, the interface degradation at the HTL/perovskite and the perovskite/ETL interfaces are accounted through interface recombination velocities ($S_v$).

Ionic effects are considered in accordance to literature[5,23–25]. Here, the positive ions ($N_{I,p}$) are immobile and spatially uniform. i.e., $N_{I,p}(x) = I_0$ ($I_0$ is a constant which denotes the average ionic density). Further, the steady state solution of the continuity equation for mobile ions is $N_{I,n} = Ke^{\frac{V(x)}{V_T}}$ ($K$ is a bias dependent prefactor and $V_T$ is the thermal voltage)[25,26]. Under ionic charge neutrality conditions, we have $\int_0^W N_{I,p} dx = \int_0^W N_{I,n} dx$. i.e., the net negative ionic charge in the active layer is the same as the net positive ionic charge ($W$ is the thickness of active layer, see Fig. 1). Extending the same with explicit usage of bias dependent spatial profile for mobile ions leads to

$$f \times I_0 \times W - \int_0^W K e^{V(x)/V_T} dx = 0 \quad (4)$$

where the parameter $f$ accounts for the imbalance in ionic charge neutrality. It is evident that eq. 4 ensures the charge neutrality of the active layer when $f = 1$ (denoted as case (i)). Deviation from such ionic charge neutrality can be due to an excess of negative mobile ions (i.e., case (ii) with $f > 1$) or due to a deficiency of mobile ions (i.e., case (iii) with $f < 1$). For ease of analysis, the net immobile ion concentration is considered as unchanged for all conditions. In addition, note that eq. (4) accounts for ionic charge imbalance in the active layer only and any electrostatic effect due to mobile ions in the TL is neglected as the TLs are, in general, heavily doped.

The modeling framework has been extensively calibrated against experimental data in our previous publications (i.e., for $f = 1$)[25–27]. Table 1 of Suppl Mat. summarizes various parameters and associated references. In addition, Section III of Suppl. Mat. shows the comparison between simulation results and experimental data (from ref. [11]).

**RESULTS AND DISCUSSIONS:**

### I.   ELECTROSTATIC ASYMMETRY

To explore the influence of ionic imbalance, we first analyze the electrostatics of the system under steady state conditions. Figures 2a-c show the energy band (EB) diagrams of the illuminated devices at near short circuit conditions. By examining the Fig. 2a-c, we observe that the ions inside the perovskite indeed screen the built-in electric field[8,28–30]. The Fermi level split inside the active layer is due to the illumination, whereas that in the transport layer is due to the over-the-barrier transport of carriers[26,31]. Note that the minority carrier concentration in the transport layers is very low (less than the corresponding intrinsic carrier density. The same effect was detailed in an earlier publication[26].

For the balanced ion condition, case (i), the $V_{BI}$ drops nearly equally at both the HTL/Perovskite and Perovskite/ETL junctions (denoted as J1 and J2 in Fig. 2a) at short circuit condition. Interestingly, asymmetrical potential barriers are observed for cases (ii) and (iii), see Figs. 2b-c. An increase in negative mobile ions in case (ii) leads to a significant band bending in the perovskite-ETL junction (see J2 in Fig. 2b). Also, as the negative mobile ions are accumulated in large quantities in the perovskite near the interface, more potential drop is seen in the ETL than in the perovskite near the interface. However, for case (iii), a prominent junction formation near the HTL-perovskite interface is seen in which the band bending extends into both the HTL and the perovskite near the interface (junction between $10^{18} cm^{-3}$ dopants of HTL and $5 \times 10^{17} cm^{-3}$ positive immobile ions of perovskite, see Fig. 2c).

The formation of asymmetric junction results in an imbalance in the carrier densities which could be interpreted, curiously, as an apparent unintentional doping of perovskite active layer. In case (i), the EB diagram indicates that the electron and hole density in the active layer is nearly the same. Here, both type of carriers are photo-generated[26]. However, for case (ii), the barrier for carrier injection is very small at J1. Similar scenario holds at J2 for case (iii). Such a reduced barrier w.r.t the TL results in contact induced carrier injection in the active layer. Accordingly, holes and electrons are the majority carrier in the active layer for case (ii) and case (iii), respectively (see Figs. 2h-i). This contact injected carriers could act as an unintentional doping[32]. Further, Mott's equation[33] can be employed to determine the average contact injected carrier density in the absorber, and is given as

$$\rho_{inj} = (2\varepsilon V_T \rho_0 e^{-\Delta\Phi_B/V_T}/qW^2)^{1/2} \quad (5)$$

Here, $\Delta\Phi_B$ is the band bending in the perovskite at the interface and $\rho_0$ is the carrier density at the interface. When $\rho_{inj}$ is higher than the excess carrier density due to generation ($\rho_{gen} = G\tau_{eff}$), then the carrier density inside the active layer would be dominated by the contact (see section IV, Suppl. Mat.).

With an increase in bias, the band bending inside the perovskite is further reduced as shown in Figs. 2d-f. For case (i), the band bending at both interfaces reduces and leads to an increase of both types of carriers inside the perovskite (see, Fig. 2g). Note that even at near open circuit conditions, case (i) has a sufficient barrier at both interfaces ($\Delta\Phi_B = 0.08eV$) which results in a $\rho_{inj} = 1 \times 10^{15}\ cm^{-3}$. As $\rho_{inj}$ is less than $\rho_{gen} = 2.8 \times 10^{15} cm^{-3}$, the carrier density is dominated by photogeneration for case (i). Section IV of Suppl. Mat contains additional details regarding this analysis. However, junction J1 for case (ii) and junction J2 for case (iii) have nearly zero barrier ($\Delta\Phi_B=0$) in the perovskite at near open circuit condition (see, Figs. e-f). This contributes to $\rho_{inj} =$ $8.7\times10^{15}cm^{-3}(>\rho_{gen})$ which results in dominant p-type (due to J1) and n-type (due to J2) behavior for case (ii) and case (iii) schemes, respectively. Accordingly, the minority carrier density in the active layer is due to photogeneration while the majority carrier density is due to contact injection from J1 for case (ii) (similar argument holds good for case iii).

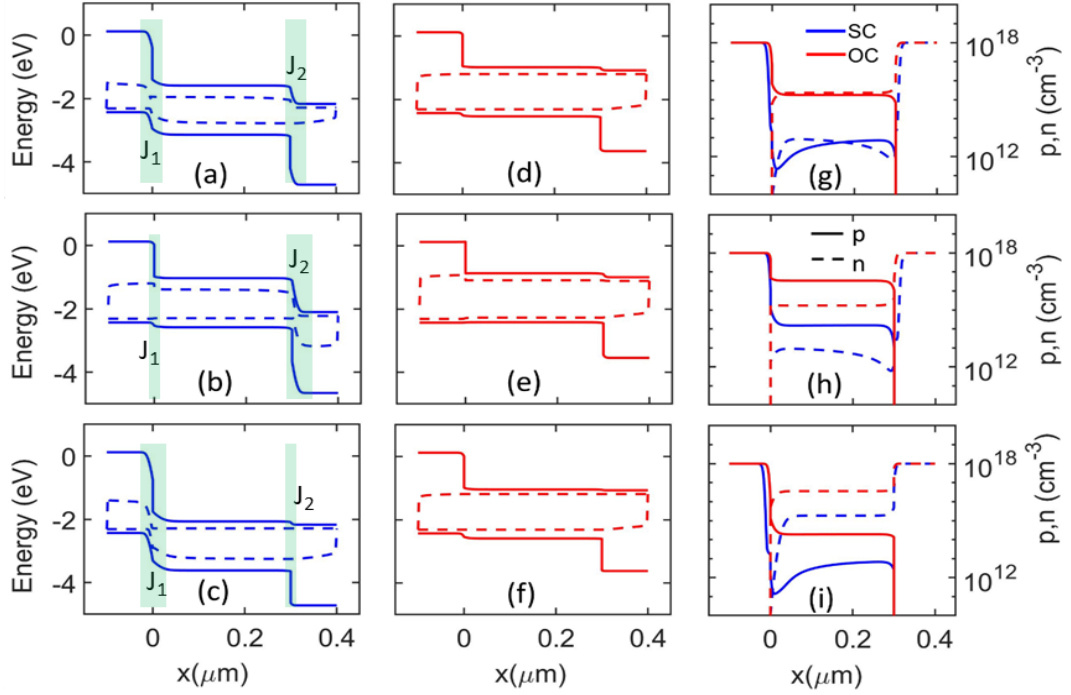

**FIG. 2**. Energy band diagram of PSC near short circuit conditions for (a) $f=1$, (b) $f>1$, (c) $f<1$). Energy band diagram of PSC near open circuit conditions for (d) $f=1$, (e) $f>1$, (f) $f<1$). Spatial profile of carrier density near short circuit and open circuit conditions for (g) $f=1$, (h) $f>1$, (i) $f<1$). Dashed lines indicate electron density and bold line indicate hole density. $G=5.21\times10^{21}\ cm^{-3}/s$ and $S_v=10^2\ cm/s$. All devices have immobile ion density of $5\times10^{17}cm^{-3}$.

The above results indicate that ion imbalance leads to an asymmetry in the device electrostatics. However, there are several factors that, otherwise, introduce asymmetry in perovskite devices. One such major aspect is the position dependent photo-induced carrier generation rate. As such, coupled with the specifics of device structure (i.e., NIP vs. PIN architecture), optimization of perovskite solar cells could be a challenging task in the presence of various non-uniformities due to carrier generation rate, ionic imbalance and variations in properties like carrier lifetime and interface recombination. Before we discuss the impact on PIN vs. NIP device architectures, the impact of ion imbalance induced asymmetrical band bending and material/interface degradation is addressed in the next section.

## II.    MATERIAL/INTERFACE DEGRADATION

Degradation of interface or bulk lifetime is known to result in performance degradation of PSCs[34–39]. The same could be influenced in a non-trivial manner due to ion imbalance in the perovskite active layer. For example, Fig. 3 shows the current-voltage characteristics of the devices with poor ($S_v=10^4\ cm/s$) and good interfaces ($S_v=10^2\ cm/s$, see inset). With better interfaces, cases (ii) and (iii) show marginally improved performance than case (i), especially at near open circuit conditions (see inset of Fig. 3). However, in the presence of significant interface degradation, the cases (ii) and (iii) show a reduced short circuit current ($J_{SC}$) and slightly improved open circuit voltage ($V_{OC}$) than the case (i). Dependence of various performance parameters on interface degradation and ion imbalance is discussed below. Along with that, the influence of spatial location of degradation and band offsets with TLs are also explored in this section.

$J_{SC}$: The variation of $J_{SC}$ with $S_v$ of the devices with different ion migration cases is shown in Fig. 4a. Clearly, a drastic decrease in $J_{SC}$ is seen with an increase in interface recombination velocity for unbalanced cases. Specifically, a three-order increase in $S_v$ causes $5\ mA/cm^2$ loss in $J_{SC}$ for case (i) whereas, a $15\ mA/cm^2$ loss of $J_{SC}$ is found for case (ii).

It is well known that high density mobile ions of screen the electric field in the active layer and hence the carrier transport becomes diffusion dominant in the PSCs[29,30]. For case (i), the band bending in the perovskite near the HTL/ETL interfaces helps the carriers overcome the interface recombination at near short circuit conditions. However, due to the asymmetric junction formation, poor band bending is seen in the perovskite near one of the junctions for cases (ii) and (iii) (see junction J1 in Fig. 2b and junction J2 in Fig. 2c). This leads to significant interface recombination of carriers at that particular interface (see section V, Suppl. Mat.). Accordingly, the effective time constant of the carriers with this one-sided interface recombination is $\frac{1}{\tau_{eff}}=\frac{1}{\tau}+\frac{S_v}{W}$. Indeed, the $J_{SC}=qGL_{eff}$ (where $L_{eff}=\sqrt{D\tau_{eff}}$ ) calculated with one side recombination well anticipates the results of case (ii) (see, dashed blue curve of Fig. 4a). As discussed, a higher positive immobile ion, as in case (iii), introduces a band bending that extends to the perovskite near the HTL interface. This aids carrier transport, and hence $J_{SC}$ reduction with interface recombination is not severe in case (iii).

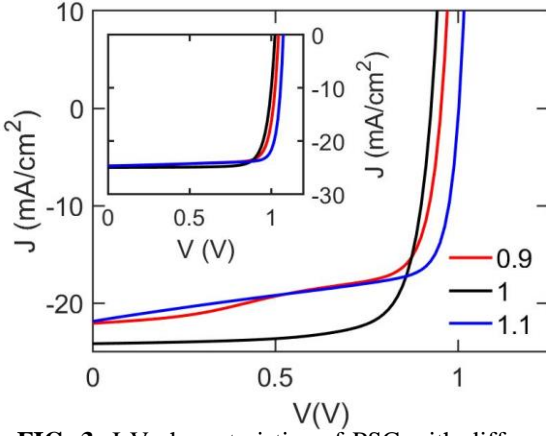

**FIG. 3**. J-V characteristics of PSC with different ion migration cases under illumination ($G = 5.21 \times 10^{21}\ cm^{-3}/s$) with $S_v = 10^4\ cm/s$. Inset shows J-V characteristics of device with $S_v = 10^2\ cm/s$. $f = 1$ indicate balanced ion case, $f = 1.1$ denotes case (ii) and $f = 0.9$ represents case (iii). All devices have immobile ion density of $5 \times 10^{17} cm^{-3}$.

$V_{OC}$: As previously discussed, at near open circuit conditions contact injection in case (i) is inhibited by the band bending at the interfaces. Hence the excess carriers inside the perovskite are only generation dependent. It is well known that the open circuit voltage ($V_{OC}$) could be written as[40–42] $V_{OC} = (\Delta E_{fn} + \Delta E_{fp})/V_T$, where $\Delta E_{fn} = (E_{fn} - E_i)$ and $\Delta E_{fp} = (E_i - E_{fp})$. Hence for $f = 1$, we have $\Delta E_{fp} = \Delta E_{fn}$ and,

$$V_{OC} = 2V_T ln(2G\tau_{eff}/n_i),\quad for\ f = 1 \quad (6)$$

Depending upon the band bending near interfaces, the $\tau_{eff}$ varies. Figure 4b shows the variation of $V_{OC}$ with $S_v$. The black bold curve indicates the trend in $V_{OC}$ of balance ion case and the black dashed line depicts the $V_{OC}$ obtained using eq. 6 (assuming one side interface recombination, $\frac{1}{\tau_{eff}} = \frac{1}{\tau} + \frac{S_v}{W}$).

As discussed earlier, the electron density of the device with case (ii) is photo-generation dependent, whereas hole density depends on contact injection ($p_{inj}$). Similarly, the devices with case (iii) have photo-generation dependent hole carrier concentration and contact injection dependent electron carrier concentration ($n_{inj}$). Hence, $\Delta E_{fp}$ for $f > 1$ and $\Delta E_{fn}$ for $f < 1$ would be $V_T ln(p_{inj}/n_i)$ and $V_T ln(n_{inj}/n_i)$, respectively. Accordingly, the $V_{OC}$ for $f \neq 1$ is

$$V_{OC} = V_T \ln(G\tau_{eff}/n_i) + V_T \ln(p_{inj}/n_i),\\ for\ f > 1 \quad (7)$$

$$V_{OC} = V_T \ln(G\tau_{eff}/n_i) + V_T \ln(n_{inj}/n_i),\\ for\ f < 1 \quad (8)$$

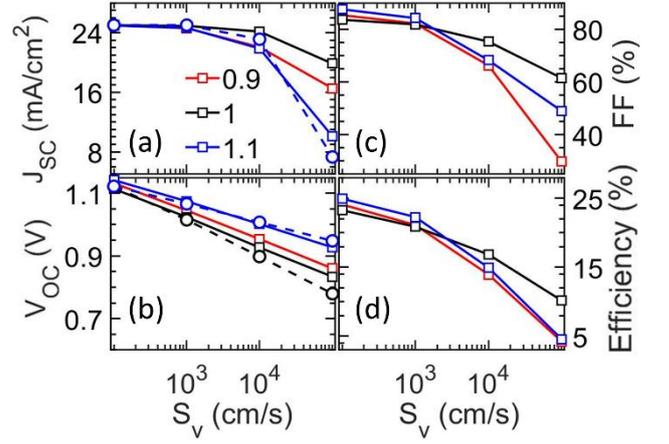

**FIG. 4**. Effect of interface recombination velocity ($S_v$) on the performance metrics of PSC for different ion migration cases (immobile ion density of $5 \times 10^{17} cm^{-3}$, $G = 5.21 \times 10^{21}\ cm^{-3}/s$).

where, $p_{inj} = (2\varepsilon V_T p_0/qW^2)^{1/2}$ and $n_{inj} = (2\varepsilon V_T n_0/qW^2)^{1/2}$ which is obtained by considering $\Delta \Phi_B = 0$ in eq. 5. Here, $p_0 = N_{dop,TL}N_{V,PER}/N_{V,TL}$ and $n_0 = N_{dop,TL}N_{C,PER}/N_{C,TL}$. The parameters $N_{C,TL}$ and $N_{V,TL}$ denote the effective density of states in the conduction band and valence band, respectively, of transport layer and $N_{dop,TL}$ is the doping density of transport layer. $N_{C,PER}$ and $N_{V,PER}$ denotes the corresponding effective density of states of perovskite.

The unintentional doping or carrier injection gives an improved performance at near open circuit condition for the imbalanced ion cases when compared to the balanced ion case. The blue and red curves in Fig. 4b show the effect of $S_v$ on $V_{OC}$ for cases (ii) and (iii) respectively. Further, the $V_{OC}$ obtained using eq. 7 is plotted as dashed blue line and is in good agreement with the case(ii). Note that, band bending in the perovskite near the HTL interface (due to excess immobile ions) restricts the buildup of carrier from injection. Hence the second component in the eq. 8 and thus the total $V_{OC}$ of case (iii) is less than that of case (ii).

**Fill Factor:** The fill factor variation with interface recombination velocity of the different ion imbalance cases is shown in Fig. 4c. An increase in $S_v$ reduces the fill factor of all three cases. However, ionic imbalance cases show much more detrimental effect. Particularly, case (iii) has poor fill factor when $S_v$ is large which is due to the formation of S-type J-V characteristics[43–45] (see section VI, Suppl. Mat.).

**Efficiency:** The effect of interface recombination velocity on the efficiency of the device is shown in Fig. 4d. For good interfaces, the device with uncompensated ions shows marginally improved performance in comparison to the device with balanced ions. This is due to the increase in $V_{OC}$ owing to contact injection. The increase in interface recombination velocity reduces $J_{SC}$ for case (ii) and FF for case (iii) which leads to the reduction in efficiency as well.

**Material degradation vs position:** In addition of interface effects, degradation of carrier lifetime in the bulk of the active layer (grain boundaries, etc.) could lead to performance loss in PSCs[3,46,47]. For this, lifetime degradation at three locations inside the active layer (a. near HTL interface, b. at mid length and c. at ETL interface) is considered. As shown in Fig. S7 (Suppl. Mat.), in all three scenarios, the impact of material degradation is greater if it occurs near the mid-length of the active layer. Further, the loss is severe in case (i) which is evident from the minority carrier profile (see Fig. 2g). However, degradation near the HTL interface and degradation near the ETL interface have a greater impact on cases (ii) and (iii), respectively. These results are expected as the recombination of carriers would be high at near J1 for case (ii) and J2 for case (iii) due to the poor band bending in the perovskite at the corresponding junctions.

**Band Offset:** The impact of ionic imbalance on device performance is influenced by the band offsets between perovskite and transport layers ($\Delta = E_{c,per} - E_{c,ETL}$ and $\Delta = E_{v,HTL} - E_{v,per}$; see section VIII, Suppl. Mat. for additional details). The offset ($\Delta$) in the energy levels create an inherent injection barrier in the device which reduces the contact injection ($\rho_{inj} < \rho_{gen}$). As a result, with good interfaces $V_{OC}$ improvement due to carrier injection is not apparent for ionic imbalance cases. However, similar to aligned energy level device ($\Delta = 0$), ionic imbalance results in a reduced $J_{SC}$ and improved $V_{OC}$ in the presence of increased interface recombination, as compared to balanced ion device (see, Fig. S8, Suppl. Mat.).

### III. NIP VS PIN SCHEME

The results discussed so far in the previous section (and Fig. S7) indicates that, under uniform carrier generation, an excess or deficiency of mobile ions could result in distinct trends w.r.t the location of degradation. This is primarily due to the asymmetry in the band bending associated with ionic imbalances. However, an inherent asymmetry exists in solar cells depending on the side of illumination – i.e., NIP vs. PIN devices. Here, we explore the combined effect of electrostatic asymmetry caused by the unbalanced ions along with the asymmetry due to illumination (i.e., NIP vs PIN scheme). For this we use position-dependent photo-generation profiles obtained through the Transfer Matrix Method[48] to analyze the effect of illumination side on JV characteristics (see Section IX, Suppl. Mat.).

For case (i), our results indicate comparable performance for both PIN and NIP schemes as they have electrostatic symmetry with significant band bending at either junction (see section VIII, Suppl. Mat.). However, results shown in Fig. 5 indicate that case (ii) shows poor performance with PIN scheme and case (iii) shows decreased current characteristics with NIP scheme. The position dependent illumination results in a higher generation of carriers in the perovskite near the illumination side compared to the other interface. And, if the band bending in the perovskite near the illumination side is poor, then the interface recombination of carriers would be higher. This leads to decrease in efficiency. Specifically, illumination near the junction J1 for case (ii) and J2 for case (iii) facilitates the interface recombination due to poor band bending (see Fig. 2 for band diagrams). On the other hand, the larger band bending helps carriers escape the interface recombination (J2 for case (ii) and J1 for case (iii)) – which explains the different trends in NIP vs. PIN schemes. Note that NIP and PIN schemes are not so distinct if the interfaces are good (see Section IX, Suppl. Mat.). Further, interestingly, Fig. 5b indicates the emergence of "double diode" behavior in J-V characteristics, a trend observed extensively in experiments[43–45], which could be caused, among other effects, by ionic imbalance and interface degradation – a possibility identified for the first time in this work.

### IV. TEMPERATURE COEFFICIENTS

The performance of solar cells under elevated temperatures is of major importance for practical purposes. Typically, the temperature dependence of solar cell is explained in terms of temperature coefficient of power conversion efficiency[49] ($T_{PCE}$, see Section X, Suppl. Mat.). To examine the same under ionic imbalance, we have considered the temperature dependence of energy band gap, mobility, effective density of states, and capture coefficients[49,50]. Figure 5c shows the trends in efficiency with temperature for all the three migration cases. As expected, an increase in temperature increases the intrinsic carrier density of the perovskite and decreases the $V_{OC}$ for all schemes[49]. This is reflected in the efficiency curve as well. Despite having a lower efficiency (or $V_{OC}$) than case (ii) at low temperatures, case (iii) shows a comparable performance at higher temperatures. For an ion density of $5 \times 10^{17} cm^{-3}$ and interface recombination velocity, $S_v = 10^2\ cm/s$, case (i) provides $T_{PCE}$ of $-0.15\ rel\%/°C$, whereas that of case (ii) is $-0.14\ rel\%/°C$, and $-0.099\ rel\%/°C$ for case (iii). These results indicate that ionic imbalance could have a positive, even if small, influence on the temperature coefficients of perovskite solar cells.

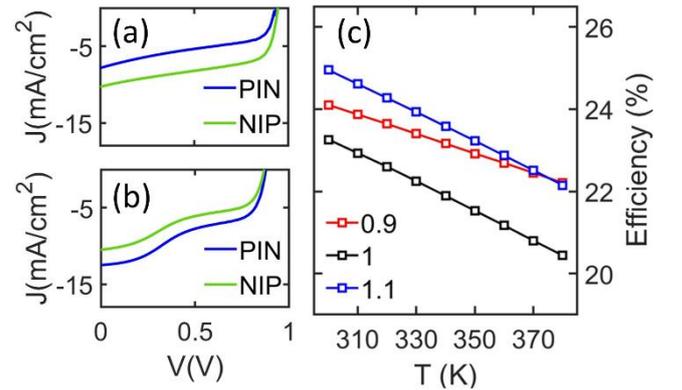

**FIG. 5**. J-V Characteristics of device with different illumination scheme for (a) $f = 1.1$, (b) $f = 0.9$ ($S_v = 5 \times 10^4\ cm/s$). (c). Effect of temperature on the efficiency of Perovskite solar cells under uniform generation ($G = 5.21 \times 10^{21}\ cm^{-3}/s$), and $S_v = 10^2\ cm/s$. Immobile ion density of $5 \times 10^{17} cm^{-3}$ is considered for all devices.

### V. HYSTERESIS

In addition to the parameters like efficiency, hysteresis indices are commonly used to assess the quality of perovskite materials along with the presence of ion migration. The influence of ionic imbalance on both parameters is elucidated in this section. It is well known in literature[5] that parameters like ionic mobility, voltage scan rate, and interface recombination influence hysteresis in perovskite devices. In accordance with literature, here our results (Fig. 6) indicate that the scan rates at which peak hysteresis occur is around 1V/s for all scenarios under consideration. We also observe that case (ii) has high peak hysteresis (hysteresis index, $HI = 0.6$) than the other two cases ($HI = 0.4$ for case (i) and $HI = 0.09$ for case (iii)). This is due to the high density of mobile ions accumulated at the interface[5,51,52] (here, case (ii) has more mobile ions) - rather, as expected, the hysteresis increases with mobile ion concentration. These simulated trends are similar to the experimentally observed trends, where extrinsic migration of $Li$ ions to perovskite leads to a high $HI$ of about 0.7 depending upon pre bias voltage[12]. Conventionally, a low HI is often associated with better interface passivation or quality. Surprisingly, our results indicate that a low HI could result due to an imbalance in ionic concentrations (as in case (iii)) – again, a possibility not yet explored in literature.

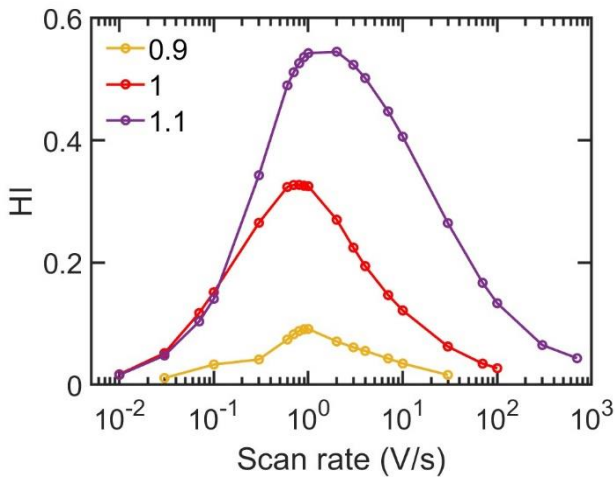

**FIG. 6.** Hysteresis vs. scan rate for different ratios of ions in the perovskite layer. All devices have immobile ion density of $4 \times 10^{17} cm^{-3}$, ion mobility of $10^{-10} cm^2 V^{-1} s^{-1}$, $S_v = 10^2 \, cm/s$, bulk SRH carrier lifetime of $\tau_{SRH} = 10^{-7} s$ and are under uniform generation, $G = 5.21 \times 10^{21} \, cm^{-3}/s$.

Our results reveal that the ionic imbalance in the active layer alters the electrostatics of the device. The asymmetry would also make PSCs more susceptible to position-dependent effects. Furthermore, ionic imbalance could be a possible cause of the apparent self-doping in the perovskite, a widely observed phenomenon[53–55]. The band bending in the transport layer due to asymmetric junction could also lead to undesired band-to-band tunnelling (see J2 for case (i), recombination between the holes in the valence band of perovskite and electrons in the conduction band of ETL) at near short-circuit conditions via a series of traps, thus degrading the performance further. Also, the unusual band bending inside transport layers for the device with ionic imbalance could prompt a criterion for transport layer thickness. Detailed discussion regarding this is provided in section XI of the supplementary material. In addition, the inter-material ion migration is greatly influenced by the ambient environment. Hence, the concerns related to ionic imbalance could be of major significance in the context of long-term stability and requires detailed experimental characterization.

## CONCLUSIONS

Using numerical simulation, we explored the effect of ionic imbalance on the performance of perovskite solar cells. Our findings suggest that PSCs with an uncompensated ionic scenario exhibit different electrostatics than devices with ion charge neutrality. For the imbalanced ion devices with good interfaces, the electrostatic asymmetry and contact injection induced apparent doping help to improve device performance including the hysteresis index. Increased interface degradation, on the other hand, reduces the efficiency of PSCs of devices with unbalanced ionic charge. The device characteristics are also greatly influenced by the illumination side and position of material degradation for devices with ionic imbalance.

## SUPPLEMENTARY MATERIAL

See the supplementary material for material parameters, derivation of carrier density from Mott equation, generation profiles using TMM, schematic of device with energetic offset, definition of TPCE.

## ACKNOWLEDGMENTS

This project is funded by Science and Engineering Research Board (SERB, project code: CRG/2019/003163), Department of Science and Technology (DST), India. Authors acknowledge IITBNF and NCPRE for computational facilities. PRN acknowledges Visvesvaraya Young Faculty Fellowship.

## DATA AVAILABILITY

The data that supports the findings of this study are available within the article [and its supplementary material].

## CORRESPONDING AUTHORS

Authors to whom correspondence should be addressed: *prnair@ee.iitb.ac.in, dhyana@iitb.ac.in*

## REFERENCES

1. Zhou W, Gu J, Yang Z, Wang M, Zhao Q. Basis and effects of ion migration on photovoltaic performance of perovskite solar cells. *J Phys D Appl Phys*. 2020;54(6). doi:10.1088/1361-6463/abbf74
2. Bertoluzzi L, Boyd CC, Rolston N, et al. Mobile Ion Concentration Measurement and Open-Access Band Diagram Simulation Platform for Halide Perovskite

2. Solar Cells. *Joule*. 2020;4(1):109-127. doi:10.1016/j.joule.2019.10.003
3. Nandal V, Nair PR. Ion induced passivation of grain boundaries in perovskite solar cells. *J Appl Phys*. 2019;125(17). doi:10.1063/1.5082967
4. Futscher MH, Lee JM, McGovern L, et al. Quantification of ion migration in CH3NH3PbI3 perovskite solar cells by transient capacitance measurements. *Mater Horiz*. 2019;6(7):1497-1503. doi:10.1039/c9mh00445a
5. Reenen S van, Kemerink M, Snaith HJ. Modeling Anomalous Hysteresis in Perovskite Solar Cells. Published online 2015. doi:10.1021/acs.jpclett.5b01645
6. Bernhardsgrütter D, Schmid MM. Modeling of Intensity-Modulated Photocurrent/Photovoltage Spectroscopy: Effect of Mobile Ions on the Dynamic Response of Perovskite Solar Cells. *Journal of Physical Chemistry C*. 2019;123(50):30077-30087. doi:10.1021/acs.jpcc.9b08457
7. Richardson G, O'Kane SEJ, Niemann RG, et al. Can slow-moving ions explain hysteresis in the current-voltage curves of perovskite solar cells? *Energy Environ Sci*. 2016;9(4):1476-1485. doi:10.1039/c5ee02740c
8. Calado P, Telford AM, Bryant D, et al. Evidence for ion migration in hybrid perovskite solar cells with minimal hysteresis. *Nat Commun*. 2016;7:1-10. doi:10.1038/ncomms13831
9. Tessler N, Vaynzof Y. Insights from Device Modeling of Perovskite Solar Cells. *ACS Energy Lett*. 2020;5(4):1260-1270. doi:10.1021/acsenergylett.0c00172
10. Ming W, Yang D, Li T, Zhang L, Du MH. Formation and Diffusion of Metal Impurities in Perovskite Solar Cell Material CH3NH3PbI3: Implications on Solar Cell Degradation and Choice of Electrode. *Advanced Science*. 2018;5(2). doi:10.1002/advs.201700662
11. Domanski K, Correa-Baena JP, Mine N, et al. Not All That Glitters Is Gold: Metal-Migration-Induced Degradation in Perovskite Solar Cells. *ACS Nano*. 2016;10(6):6306-6314. doi:10.1021/acsnano.6b02613
12. Li Z, Xiao C, Yang Y, et al. Extrinsic ion migration in perovskite solar cells. *Energy Environ Sci*. 2017;10(5):1234-1242. doi:10.1039/c7ee00358g
13. Cacovich, S., Cinà, L., Matteocci, F., Divitini, G., Midgley, P. A., Di Carlo, A., & Ducati C. Gold and iodine diffusion in large area perovskite solar cells under illumination. *Nanoscale*. 2017;9:4700-4706.
14. Kerner RA, Heo S, Roh K, MacMillan K, Larson BW, Rand BP. Organic Hole Transport Material Ionization Potential Dictates Diffusion Kinetics of Iodine Species in Halide Perovskite Devices. *ACS Energy Lett*. 2021;6(2):501-508. doi:10.1021/acsenergylett.0c02495
15. Kerner RA, Zhao L, Harvey SP, Berry JJ, Schwartz J, Rand BP. Low Threshold Voltages Electrochemically Drive Gold Migration in Halide Perovskite Devices. *ACS Energy Lett*. 2020;5(11):3352-3356. doi:10.1021/acsenergylett.0c01805
16. Ginting RT, Jeon MK, Lee KJ, Jin WY, Kim TW, Kang JW. Degradation mechanism of planar-perovskite solar cells: correlating evolution of iodine distribution and photocurrent hysteresis. *J Mater Chem A Mater*. 2017;5(9):4527-4534. doi:10.1039/c6ta09202k
17. Boyd CC, Cheacharoen R, Leijtens T, McGehee MD. Understanding Degradation Mechanisms and Improving Stability of Perovskite Photovoltaics. *Chem Rev*. 2019;119(5):3418-3451. doi:10.1021/acs.chemrev.8b00336
18. Boyd CC, Cheacharoen R, Bush KA, Prasanna R, Leijtens T, McGehee MD. Barrier Design to Prevent Metal-Induced Degradation and Improve Thermal Stability in Perovskite Solar Cells. *ACS Energy Lett*. 2018;3(7):1772-1778. doi:10.1021/acsenergylett.8b00926
19. Siegler TD, Dunlap-Shohl WA, Meng Y, et al. Water-Accelerated Photooxidation of CH$_3$NH$_3$PbI$_3$ Perovskite. *J Am Chem Soc*. Published online 2022. doi:10.1021/jacs.2c00391
20. Scharfetter, D. L., & Gummel HK. Large-signal analysis of a silicon read diode oscillator. *IEEE Trans Electron Devices*. 1969;16(1):64-77.
21. Agarwal S, Nair PR. Device engineering of perovskite solar cells to achieve near ideal efficiency. *Appl Phys Lett*. 2015;107(12). doi:10.1063/1.4931130
22. Agarwal S, Nair PR. Performance loss analysis and design space optimization of perovskite solar cells. *J Appl Phys*. 2018;124(18). doi:10.1063/1.5047841
23. Rivkin B, Fassl P, Sun Q, Taylor AD, Chen Z, Vaynzof Y. Effect of Ion Migration-Induced Electrode Degradation on the Operational Stability of Perovskite Solar Cells. *ACS Omega*. 2018;3(8):10042-10047. doi:10.1021/acsomega.8b01626
24. de Bastiani M, Dell'Erba G, Gandini M, et al. Ion migration and the role of preconditioning cycles in the stabilization of the J-V characteristics of inverted hybrid perovskite solar cells. *Adv Energy Mater*. 2016;6(2):1-9. doi:10.1002/aenm.201501453
25. Saketh Chandra T, Singareddy A, Hossain K, et al. Ion mobility independent large signal switching of perovskite devices. *Appl Phys Lett*. 2021;119(2):1-5. doi:10.1063/5.0051342
26. Sivadas D, Bhatia S, Nair PR. Efficiency limits of perovskite solar cells with n-type hole extraction layers. *Appl Phys Lett*. 2021;119(20). doi:10.1063/5.0059221
27. Singareddy A, Sadula UKR, Nair PR. Phase segregation induced efficiency degradation and variability in mixed halide perovskite solar cells. *J Appl Phys*. 2021;130(22). doi:10.1063/5.0062818
28. Bertoluzzi L, Belisle RA, Bush KA, Cheacharoen R, Mcgehee MD, Regan BCO. In Situ Measurement of Electric-Field Screening in Hysteresis-Free. Published online 2018. doi:10.1021/jacs.8b04405
29. Lopez-Varo P, Jiménez-Tejada JA, García-Rosell M, et al. Effects of Ion Distributions on Charge Collection in Perovskite Solar Cells. *ACS Energy*

Lett. 2017;2(6):1450-1453. doi:10.1021/acsenergylett.7b00424
30. Pockett A, Carnie MJ. Ionic Influences on Recombination in Perovskite Solar Cells. *ACS Energy Lett*. 2017;2(7):1683-1689. doi:10.1021/acsenergylett.7b00490
31. Pierret RF. *Semiconductor Device Fundamentals*.; 1996.
32. Nigam A, Premaratne M, Nair PR. On the validity of unintentional doping densities extracted using Mott-Schottky analysis for thin film organic devices. *Org Electron*. 2013;14(11):2902-2907. doi:10.1016/j.orgel.2013.08.005
33. Mott, Francis N, Gurney RW. *Electronic Processes in Ionic Crystals*. Clarendon Press; 1948.
34. Wang C, Zhang C, Huang Y, et al. Degradation behavior of planar heterojunction CH3NH3PbI3 perovskite solar cells. *Synth Met*. 2017;227:43-51. doi:10.1016/j.synthmet.2017.02.022
35. Shao S, Loi MA. The Role of the Interfaces in Perovskite Solar Cells. *Adv Mater Interfaces*. 2020;7(1). doi:10.1002/admi.201901469
36. Peng W, Anand B, Liu L, et al. Influence of growth temperature on bulk and surface defects in hybrid lead halide perovskite films. *Nanoscale*. 2016;8(3):1627-1634. doi:10.1039/c5nr06222e
37. Wu J, Shi J, Li Y, et al. Quantifying the Interface Defect for the Stability Origin of Perovskite Solar Cells. *Adv Energy Mater*. 2019;9(37). doi:10.1002/aenm.201901352
38. Xiong J, Yang B, Cao C, et al. Interface degradation of perovskite solar cells and its modification using an annealing-free TiO2 NPs layer. *Org Electron*. 2016;30:30-35. doi:10.1016/j.orgel.2015.12.010
39. Lee H, Lee C. Analysis of Ion-Diffusion-Induced Interface Degradation in Inverted Perovskite Solar Cells via Restoration of the Ag Electrode. *Adv Energy Mater*. 2018;8(11). doi:10.1002/aenm.201702197
40. Sinton RA, Cuevas A. Contactless determination of current-voltage characteristics and minority-carrier lifetimes in semiconductors from quasi-steady-state photoconductance data. *Appl Phys Lett*. 1996;69(17):2510-2512. doi:10.1063/1.117723
41. Nandal V, Nair PR. Predictive Modeling of Ion Migration Induced Degradation in Perovskite Solar Cells. *ACS Nano*. 2017;11(11):11505-11512. doi:10.1021/acsnano.7b06294
42. Agarwal S, Seetharaman M, Kumawat NK, et al. On the uniqueness of ideality factor and voltage exponent of perovskite-based solar cells. *Journal of Physical Chemistry Letters*. 2014;5(23):4115-4121. doi:10.1021/jz5021636
43. Guerrero A, You J, Aranda C, et al. Interfacial degradation of planar lead halide perovskite solar cells. *ACS Nano*. 2016;10(1):218-224. doi:10.1021/acsnano.5b03687
44. Lin Z, Chang J, Xiao J, et al. Interface studies of the planar heterojunction perovskite solar cells. *Solar Energy Materials and Solar Cells*. 2016;157:783-790. doi:10.1016/j.solmat.2016.07.045
45. Ghosh S, Singh R, Subbiah AS, Boix PP, Mora Seró I, Sarkar SK. Enhanced operational stability through interfacial modification by active encapsulation of perovskite solar cells. *Appl Phys Lett*. 2020;116(11). doi:10.1063/1.5144038
46. Li D, Bretschneider SA, Bergmann VW, et al. Humidity-Induced Grain Boundaries in MAPbI3 Perovskite Films. *Journal of Physical Chemistry C*. 2016;120(12):6363-6368. doi:10.1021/acs.jpcc.6b00335
47. deQuilettes DW, Vorpahl S, D. Stranks S, et al. Impact of microstructure on local carrier lifetime in perovskite solar cells. *Science (1979)*. 2015;348(6235):683-686. doi:10.1126/science.aaa4157
48. Pettersson LAA, Roman LS, Inganäs O. Modeling photocurrent action spectra of photovoltaic devices based on organic thin films. *J Appl Phys*. 1999;86(1):487-496. doi:10.1063/1.370757
49. Moot T, Patel JB, McAndrews G, et al. Temperature Coefficients of Perovskite Photovoltaics for Energy Yield Calculations. *ACS Energy Lett*. 2021;6(5):2038-2047. doi:10.1021/acsenergylett.1c00748
50. Aydin E, Allen TG, de Bastiani M, et al. Interplay between temperature and bandgap energies on the outdoor performance of perovskite/silicon tandem solar cells. *Nat Energy*. 2020;5(11):851-859. doi:10.1038/s41560-020-00687-4
51. Li C, Tscheuschner S, Paulus F, et al. Iodine Migration and its Effect on Hysteresis in Perovskite Solar Cells. *Advanced Materials*. 2016;28(12):2446-2454. doi:10.1002/adma.201503832
52. Tress W, Marinova N, Moehl T, Zakeeruddin SM, Nazeeruddin MK, Grätzel M. Understanding the rate-dependent J-V hysteresis, slow time component, and aging in CH3NH3PbI3 perovskite solar cells: The role of a compensated electric field. *Energy Environ Sci*. 2015;8(3):995-1004. doi:10.1039/c4ee03664f
53. Peng W, Yin J, Ho KT, et al. Ultralow Self-Doping in Two-dimensional Hybrid Perovskite Single Crystals. *Nano Lett*. 2017;17(8):4759-4767. doi:10.1021/acs.nanolett.7b01475
54. Shi T, Yin WJ, Hong F, Zhu K, Yan Y. Unipolar self-doping behavior in perovskite CH3NH3PbBr3. *Appl Phys Lett*. 2015;106(10). doi:10.1063/1.4914544
55. Cho SH, Byeon J, Jeong K, et al. Investigation of Defect-Tolerant Perovskite Solar Cells with Long-Term Stability via Controlling the Self-Doping Effect. *Adv Energy Mater*. 2021;11(17). doi:10.1002/aenm.202100555